\documentclass[11pt,letterpaper,english,aps,manuscript]{revtex4}
\usepackage{fourier}
\usepackage{avant}
\usepackage[T1]{fontenc}
\usepackage[latin9]{inputenc}
\usepackage{amsmath}
\usepackage{graphicx}
\usepackage{amssymb}
\usepackage{esint}

\makeatletter

\pdfpageheight\paperheight
\pdfpagewidth\paperwidth

\@ifundefined{textcolor}{}
{%
 \definecolor{BLACK}{gray}{0}
 \definecolor{WHITE}{gray}{1}
 \definecolor{RED}{rgb}{1,0,0}
 \definecolor{GREEN}{rgb}{0,1,0}
 \definecolor{BLUE}{rgb}{0,0,1}
 \definecolor{CYAN}{cmyk}{1,0,0,0}
 \definecolor{MAGENTA}{cmyk}{0,1,0,0}
 \definecolor{YELLOW}{cmyk}{0,0,1,0}
 }

\makeatother

\usepackage{babel}

\begin{document}

\title{Recurrent host mobility in spatial epidemics: beyond reaction-diffusion}

\author{Vitaly Belik }

\affiliation{Max Planck Institute for Dynamics and Self-Organization, G\"ottingen,
Germany}

\affiliation{Department of Civil and Environmental Engineering, Massachusetts
Institute of Technology, Cambridge, MA, USA}

\author{Theo Geisel}

\affiliation{Max Planck Institute for Dynamics and Self-Organization, G\"ottingen,
Germany}

\affiliation{Department of Physics, University of G\"ottingen, G\"ottingen,
Germany}

\author{Dirk Brockmann}

\affiliation{Northwestern Institute on Complex Systems}

\affiliation{Department of Engineering Sciences and Applied Mathematics, Northwestern
University, Evanston, IL, USA}
\begin{abstract}
Human mobility is a key factor in spatial disease dynamics and related
phenomena. In computational models host mobility is typically modeled
by diffusion in space or on metapolulation networks. Alternatively,
an effective force of infection across distance has been introduced
to capture spatial dispersal implicitly. Both approaches do not account
for important aspects of natural human mobility, diffusion does not
capture the high degree of predictability in natural human mobility
patters, e.g. the high percentage of return movements to individuals'
base location, the effective force of infection approach assumes immediate
equilibrium with respect to dispersal. These conditions are typically
not met in natural scenarios. We investigate an epidemiological model
that explicitly captures natural individual mobility patterns. We
systematically investigate generic dynamical features of the model
on regular lattices as well as metapopulation networks and show that
generally the model exhibits significant dynamical differences in
comparison to ordinary diffusion and effective force of infection
models. For instance, the natural human mobility model exhibits a
saturation of wave front speeds and a novel type of invasion threshold
that is a function of the return rate in mobility patterns. In the
light of these new findings and with the availability of precise and
pervasive data on human mobility our approach provides a framework
for a more sophisticated modeling of spatial disease dynamics.
\end{abstract}
\maketitle

\section{Introduction}

\begin{figure}[th]
\includegraphics[width=1\columnwidth]{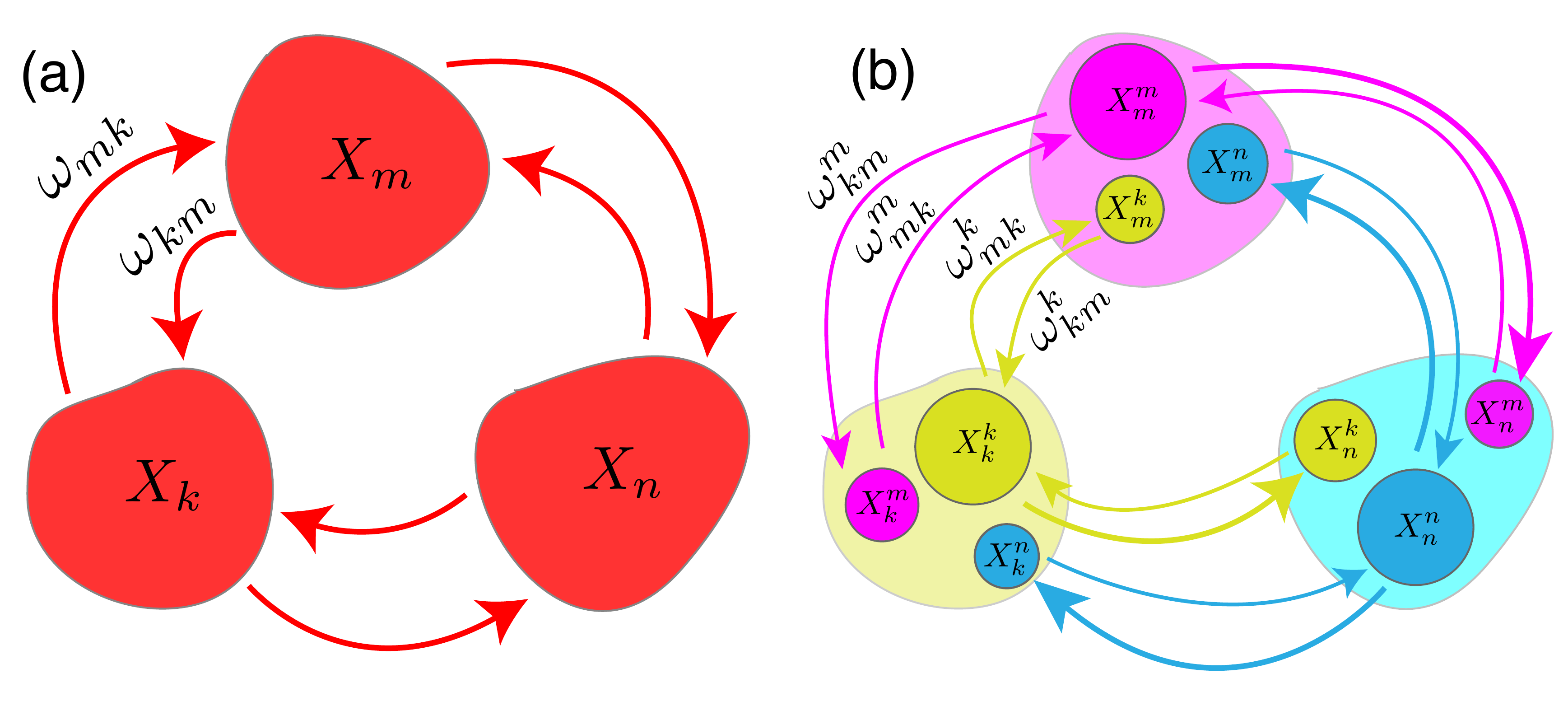}

\caption{\textbf{Models for human mobility:} Patches and arrows represent individual
populations and travel flux, respectively. (a) \emph{Diffusive dispersal:}
indistinguishable individuals travel randomly between different locations
governed by the set of transition rates $\omega_{nm}$. (b) \emph{Natural
human mobility:} Individuals labeled $k$ travel from their base location
$k$ to connected locations $m$ and back with travel rates $\omega_{mk}^{k}$
and $\omega_{km}^{k}$, respectively.\label{fig:mechanism}}

\end{figure}

The 2009 outbreak of a novel subtype (H1N1) of influenza A and its
subsequent worldwide spread, the recent emergence of new human infectious
diseases such as SARS in 2003, and the recurrent seasonal ourbreaks
of influenza epidemics illustrate the growing importance of understanding
the dynamics of human infectious diseases~\cite{may,HBGPnas,Fraser:2009ek}.
Key to understanding spatial dynamics in particular is an accurate
assessments of human mobility patterns as infectious diseases spread
among different locations due to movements of their host. Despite
recent advances~\cite{DBNature,Gonzalez2008} comprehensive data
on mobility is typically unavailable, and modelers have to make reasonable
assumptions when implementing host mobility in models. Often it is
assumed that hosts move randomly (Fig.\ref{fig:mechanism}(a)) in
the system yielding reaction-diffusion dynamics~\cite{Rileyreview,HBGPnas,Brockmann2007,kpp,fisher,rvachev,Vesp1}.
An alternative heuristic approach captures spatial dynamics without
explicitely accounting for host dispersal~\cite{Rushton1955,hagernaas}.
Instead an effective force of infection between spatially separated
populations mimics the effect of disease transmission across distance.
Typically, this force is assumed to be proportional to the prevalence
of the disease in one of the locations. Yet, because this approach
lacks the explicit incorporation of host movements, a systematic analysis
of conditions under which it is applicable is difficult.

Recently, human movement patterns became accessible based on pervasive
mobility proxies~\cite{DBNature,Gonzalez2008,Belik:2009fi,Brockmann:2009tf,Chaoming-Song-and-Zehui-Qu-and-Nicholas-Blumm-and-Albert-Laszlo-Barabasi-:2010la,Song:2010kx}.
One of the key findings of these studies confirmed the intuitive notion
that humans spend most of their time in small sets of particular locations
(home, work, shopping sites, etc.) and a person's mobility occurs
predominantly between these individual-specific locations. Furthermore,
a typical characteristic is the existence of one or two locations
that function as an individuals base location, e.g. their homes to
which individuals typically return before they travel to another place.
A key feature of human mobility therefore is a bi-directional pattern
in their trajectories among small sets of salient locations contrasting
diffusion processes in which individual agents eventually visit every
location in the entire system. Topologically, natural human mobility
patterns can be described by individual mobility networks that possess
a hub-and-spokes structure, in which a central hub represents a base
location and a limited set of places connected by spokes the set of
popular destination locations. Spreading phenomena across a large
spatial scale occurs by virtue of interactions of agents that possess
overlapping individual mobility networks. 

In preceding papers and in this conference we have presented a stochastic
model that explicitly accounts for the natural human mobility patterns
described above~\cite{Belik:2009fi,Belik:2011}. In particular, it
respects the fact that individuals typically return to their unique
base location before they travel to a new destination (recently a
similar approach has been used by Balcan and Vespignani~\cite{Balcan:2011gv}).
Here we investigate properties of this model and demonstrate that
the dynamic consequences of natural mobility patterns are profound.
Its basic aspects are depicted in Fig.~\ref{fig:mechanism}. In this
model mobility of the entire population is represented by a set of
overlapping individual mobility networks. In the language of complex
network theory each individual mobility pattern consists of a central
node (the base location) connected to a set of accessible destinations
(connected nodes) in the aforementioned hub-spokes topology. 

Although mathematical metapopulation models have been proposed that
are able to capture natural human mobility patterns~\cite{bidirect1,Keeling2002}
it has remained elusive to what extent and under which conditions
such models exhibit dynamic features that are qualitatively different
from ordinary reaction-diffusion processes. It is unclear how these
systems can be related to paradigmatic reaction-diffusion systems. 

In the present article we show that the dynamics exhibits profound
differences, indeed, as compared to ordinary reaction-diffusion systems.
We concentrate on the analysis of epidemics on regular lattice and
complex metapopulation networks. On lattices we obtain a generalization
of the paradigmatic Fisher-Kolmogorov equation that describes wave
propagation in reaction-diffusion systems. We show that the spatially
continuous version of our model exhibits travelling wave solutions
and compute their front velocities as a function of system parameters.
Contrary to reaction-diffusion systems that exhibit a monotonic and
unbounded increase of the front velocity with increasing travel rate,
our model predicts an upper bound for front velocities. We show that
the front shape strikingly differs from those predicted by reaction-diffusion
systems and is more robust in response to changes in parameters. We
introduce a commuting ratio parameter, a quantity present only in
the natural mobility model, and investigate the front velocity as
a function of it. Analysing a fully stochastic system in regular lattices
as well as complex metapopulation networks we find that a global outbreak
of a disease is determined by a novel threshold that is determined
by the typical time spent away from individuals' base locations.

\section{Natural Human Mobility and Disease Dynamics}

\subsection{Disease Dynamics on a Metapopulation}

We consider a system of populations labeled $m=1,...,M$ and assume
that in each population an epidemic outbreak can be described by a
compartmental SIR-model, i.e.\begin{equation}
I_{m}+S_{m}\xrightarrow{\alpha}2I_{m},\quad I_{m}\xrightarrow{\beta}R_{m},\label{eq:reaction}\end{equation}
in which the reactions govern infection due to the interaction of
infected ($I$) with susceptible ($S$) individuals at rate $\alpha$,
and recovery of an infected individuals at rate $\beta$, respectively.
The number of individuals in a population is given by $N_{m}=S_{m}+I_{m}+R_{m}$.
The spread of an epidemic across the set of $M$ populations is governed
by the exchange of individuals between populations. The most prominent
and conceptually clearest ansatz is diffusive dispersal between populations
in which individuals of each class move from location $n$ to $m$
at rate $\omega_{mn}$, i.e \begin{equation}
X_{n}\xrightarrow{\omega_{mn}}X_{m},\label{eq:diffusion}\end{equation}
where $X_{n}$ represents $I_{n}$, $S_{n}$ or $R_{n}$. The rates
$\omega_{mn}$ generate an equilibrium distribution $\overline{N}_{n}$
of individuals among $M$ populations. Assuming that for a pair of
populations exchange rates are nonzero, detailed balance is fullfilled,
i.e. $\overline{N}_{n}/\overline{N}_{m}=\omega_{nm}/\omega_{mn}$.
In the following we will assume that the system is in equilibrium
with respect to dispersal, i.e. $N_{m}=\overline{N}_{m}$, yielding
the following mean-field dynamical equations: \begin{eqnarray}
\partial_{t}I_{n} & = & \alpha S_{n}I_{n}/\overline{N}_{n}-\beta I_{n}+\sum_{m\neq n}\left(\omega_{nm}I_{m}-\omega_{mn}I_{n}\right)\nonumber \\
\partial_{t}S_{n} & = & -\alpha S_{n}I_{n}/\overline{N}_{n}+\sum_{m\neq n}\left(\omega_{nm}S_{m}-\omega_{mn}S_{n}\right)\nonumber \\
R_{n} & = & \overline{N}_{n}-S_{n}-I_{n}.\label{eq:RDNetworkModel}\end{eqnarray}
Note that in the reaction-diffusion system individuals are indistinguishable
apart from their infection status and move about randomly between
all available locations $m$. This approach has been employed both
in complex networks of coupled populations as well as simplified lattice
models~\cite{CollizaVespignani2007,Vesp1}. 

The relation to spatially continuous models is best illustrated in
a system of linearly aligned populations separated by distance $l$,
at locations $x=nl$, uniform population size $\overline{N}_{n}=\overline{N}=\text{const}.$
and travel between neighboring populations only, i.e. $\omega_{nm}=\omega\delta_{n-1,m}+\omega\delta_{n+1,m}$.
The overall uniform rate $\omega$ is related to the waiting time
in a given location $\tau=\omega^{-1}.$ In the limit $l,\tau\rightarrow0$
with $D=l^{2}/\tau$ this model yields the 1-d reaction-diffusion
system\begin{eqnarray}
\partial_{t}j & = & \alpha js-\beta j+D\partial_{x}^{2}j\nonumber \\
\partial_{t}s & = & -\alpha js+D\partial_{x}^{2}s,\label{eq:FK_kind_of}\end{eqnarray}
where $j(x,t)l=I_{n}/\overline{N}_{n}$ and $s(x,t)l=S_{n}/\overline{N}_{n}$.
These equations are related to the Fisher-Kolmogorov equation~\cite{fisher,kpp}.
For sufficiently localized initial conditions $j(x,t=0)$ this system
exhibits travelling waves with front velocity \begin{equation}
c=2l\sqrt{(\alpha-\beta)/\tau}\sim\sqrt{\omega},\label{eq:c_FK}\end{equation}
 that monotonically increases with the global mobility rate $\omega$. 

In order to account for individual mobility networks that exhibit
base locations and natural recurrent movements we propose the following
generalization of Eqs.~\eqref{eq:reaction},\eqref{eq:diffusion}:
We assume that individuals can be grouped into subpopulations defined
by two indices, $n$ and $k$. The first index determines the current
location $n$, the second the base location $k$. Generally, the dispersal
dynamics is then governed by a set of reactions:\begin{equation}
X_{n}^{k}\overset{\omega_{mn}^{k}}{\underset{\omega_{nm}^{k}}{\rightleftharpoons}}X_{m}^{k}.\label{eq:bronkhorst}\end{equation}
This implies that individuals of class $k$ possess their specific
dispersal rate matrix $\omega_{mn}^{k}$ that is conditioned on the
base location $k$. The rate $\omega_{mn}^{k}$ determines how individuals
of type $k$ travel from location $n$ to $m$, for fixed $k$ the
matrix $\omega_{mn}^{k}$ represents the aforementioned individual
mobility network for individuals of type $k$. The dynamical system,
incorporating disease dynamics, is given by\begin{eqnarray}
\partial_{t}I_{n}^{k} & = & \frac{\alpha}{N_{n}}S_{n}^{k}\sum_{m}I_{n}^{m}-\beta I_{n}^{k}+\sum_{m}\left(\omega_{nm}^{k}I_{m}^{k}-\omega_{mn}^{k}I_{m}^{k}\right)\nonumber \\
\partial_{t}S_{n}^{k} & =- & \frac{\alpha}{N_{n}}S_{n}^{k}\sum_{m}I_{n}^{m}+\sum_{m}\left(\omega_{nm}^{k}S_{m}^{k}-\omega_{mn}^{k}S_{m}^{k}\right),\label{eq:Bidi}\end{eqnarray}
where $I_{n}^{k}$ and $S_{n}^{k}$ are the number of infecteds and
of susceptibles of type $k$ located at $n$, respectively. $N_{n}$
denotes the total number of individuals in location $n$, i.e. $N_{n}=\sum_{k}(I_{n}^{k}+S_{n}^{k}+R_{n}^{k})$.
Note that if the rates $\omega_{nm}^{k}$ are independent of $k$,
we recover the ordinary reaction-diffusion case. 

In the following we consider the case of overlapping hub-spokes networks
corresponding to commuting between base and destination locations.
This imposes restrictions on the rates $\omega_{nm}^{k}$, it implies
that $\omega_{nm}^{k}=0$ if $k\neq n$ and $k\neq m$. That means
individuals of type $k$ that are located at $m$ cannot travel to
$n$ without returning to $k$ first. We further assume that $\omega_{mk}^{k}=\omega^{-}$,
i.e. the return rate is uniform for all $k$ and $m$. This assumption
implies that individuals typically spent the same amount of time in
distant locations before returning to their base. Assuming that travel
of the entire system is equilibrated, we obtain\[
\overline{N}_{n}=\sum_{k}\frac{\delta_{nk}+(1-\delta_{nk})\omega_{kn}^{k}/\omega_{nk}^{k}}{1+\sum_{m\neq k}\omega_{mk}^{k}/\omega_{km}^{k}}N^{k},\]
where $\overline{N}_{n}$ is the stationary number of individuals
located in population $n$ and $N^{k}=\sum_{n}N_{n}^{k}$ is the total
number of individuals of type $k$ (i.e. they belong to base location
$k$). 

An important limiting case is a situation in which mobility rates
are large compared to the rates associated with the infection and
recovery dynamics, i.e. $\omega_{mk}^{k},\omega^{-}\gg\alpha,\beta$.
In this case detailed balance is fulfilled for infecteds and susceptibles
separately and the last terms in Eq.~\eqref{eq:Bidi} vanish. If
we assume that $\omega_{mk}^{k}/\omega^{-}\ll1$ which implies that
individuals belonging to $k$ remain at their base most of time, Eq.~\eqref{eq:Bidi}
can be reduced to the effective force of infection model~\cite{Rushton1955}:
\begin{equation}
\frac{d}{dt}I^{k}=\alpha S^{k}\sum_{m}\epsilon_{km}I^{m}-\beta I^{k},\label{eq:brodel}\end{equation}
where $I^{k}=\sum_{m}I_{m}^{k}$ is the number of infected individuals
belonging to location $k$ and coupling strengths $\epsilon_{nk}=\sum_{m}p_{n}^{m}p_{k}^{m}/\overline{N}^{m}$
are explicitly related to travel rates and $p_{n}^{m}=\overline{N}_{n}^{m}/\overline{N}_{n}$
is the occupation probability. Hence direct coupling represents a
special case of our model, see also~\cite{Keeling2002}.

In order to investigate the dynamic consequences of natural human
mobility patterns as captured by Eqs.~(\ref{eq:Bidi}) ,we consider
a system analogous to the one-dimensional spatially homogeneous system
leading to the reaction-diffusion Eq.~\eqref{eq:FK_kind_of}. We
consider a 1-d lattice of populations of size $N$ separated by a
distance $l$, assume next-neighbor coupling and allow only infecteds
to travel (relaxing this restriction does not change the main results
but eases the analysis). We denote the number of infecteds at their
base location by $I_{n}^{n}$ and the number of infecteds at neighboring
locations $(n-1)$ and $(n+1)$ by $I_{n}^{-}$ or $I_{n}^{+}$, respectively.
This yields\begin{eqnarray}
I_{n}^{n}+S_{n} & \stackrel{\alpha}{\rightarrow} & 2I_{n}^{n}\nonumber \\
I_{n\mp1}^{\pm}+S_{n} & \stackrel{\alpha}{\rightarrow} & I_{n\mp1}^{\pm}+I_{n}^{n}\nonumber \\
I_{n}^{n} & \overset{\omega^{+}}{\underset{\omega^{-}}{\rightleftharpoons}} & I_{n}^{\pm}\label{eq:wurst}\end{eqnarray}
where $S_{n}$ denotes the number of susceptibles at $n$. $\omega^{+}$
and $\omega^{-}$ denote forward and return rates, respectively. In
the corresponding dynamical system we can approximate $S_{n},I_{n}^{\pm}$
by their continuous counterparts: $I_{n\pm1}^{\pm}\rightarrow I^{\pm}(x\pm l)\approx I^{\pm}\pm l\nabla I^{\pm}+\frac{l^{2}}{2}\Delta I^{\pm}$.
In an equilibrated homogeneous lattice the size of a single population
remains constant during an epidemic. Introducing concentrations $u_{n}=I_{n}/N$,
$v_{n}=(I_{n}^{+}+I_{n}^{-})/N$ and $w_{n}=(I_{n}^{+}-I_{n}^{-})/N$
this yields: \begin{eqnarray}
\partial_{t}u & = & \alpha(1-u-v)(u+v+D\Delta v+l\nabla w)+\omega^{-}v-2\omega^{+}u\nonumber \\
\partial_{t}v & = & 2\omega^{+}u-\omega^{-}v\label{eq:uv}\\
\partial_{t}w & = & -\omega^{-}w,\nonumber \end{eqnarray}
where $D=l^{2}/2$. The third equation is solved by $w\sim e^{-\omega^{-}t}$
and for $t\gg1/\omega^{-}$ yields $w\approx0$. We can therefore
discard $w$ leaving only the first two equations in~\eqref{eq:uv}.
Steady states are $\overline{u}=0$, $\overline{v}=0$ and $\overline{u}=\omega^{-}/(2\omega^{+}+\omega^{-})$,
$\overline{v}=2\omega^{+}/(2\omega^{+}+\omega^{-})$. In the non-zero
steady state concentration of infecteds in one city is given by $u+v=1$.
In order to compare this system to the reaction-diffusion model, we
calibrate both systems such as to keep the total flux of individuals
between two particular locations equal in both systems.

\section{Results}

\begin{figure}[t]
\hfill{}\includegraphics[width=0.49\columnwidth]{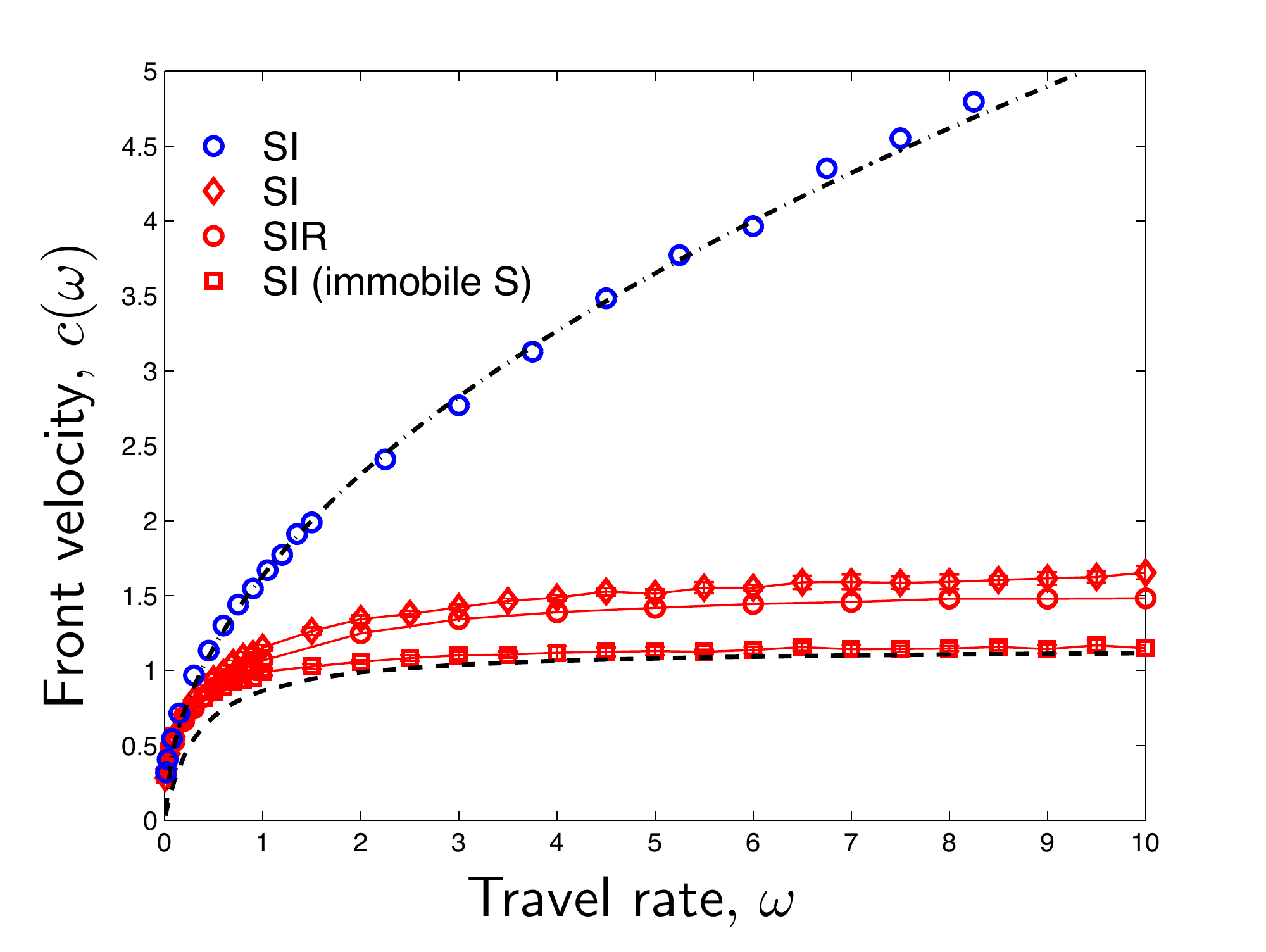}\hfill{}\includegraphics[width=0.49\columnwidth]{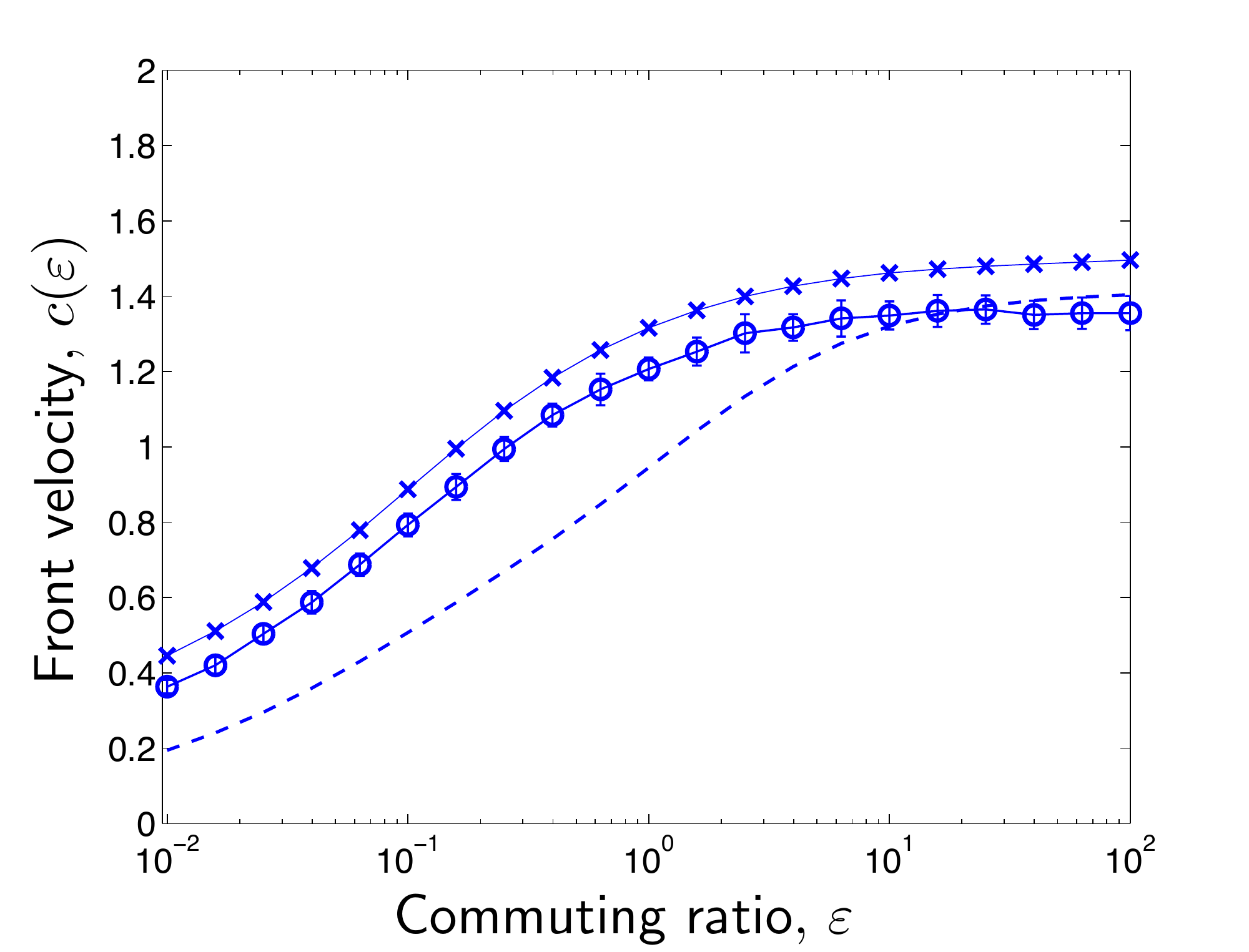}\hfill{}

\caption{\textbf{Front propagation in natural human mobility models.} \emph{Left:}
Front velocity $c(\omega)$ as a function of travel rate $\omega$
for a reaction-diffusion and the natural mobility model. Numerical
results of the stochastic simulations with $N=10^{4}$ agents per
site are depicted by symbols for reaction-diffusion (blue) and natural
mobility (red) models. Analytical results for the reaction-diffusion
system, Eqs.~\eqref{eq:c_FK} and natural mobility model, Eqs.~\eqref{eq:MV}
are depicted by dash-dotted and dashed lines, respectively. \emph{Right:}
Front velocity $c(\varepsilon)$ as a function of the commuter ratio
$\varepsilon$ (for $\omega=1$). The dashed curve illustrates the
analytical result, Eq.~\eqref{eq:Vwae}. Crosses and circles depict
results of numerical solutions of the mean-field equations \eqref{eq:Bidi}
and stochastic simulations of the natural mobility in a spatial SI
model, respectively. \label{fig:frontspeeds}}

\end{figure}

In epidemiology key questions concern conditions under which an epidemic
can spread. In this case it is generally the first task to compute
or estimate the speed at which an epidemic proliferates throughout
the entire system. On regular lattices with dispersal only among adjacent
site this task is equivalent to computing asymptotic wavefront speeds.
In complex network topologies alternative quantities are useful, e.g.
the time to reach the epidemic peak. Below, we investigate the velocity
of epidemic propagation on a lattice, and subsequently provide results
in a complex network topology.

\subsection{Front velocity in lattice systems}

Using the traveling wave ansatz $f(x,t)=g(x-ct)$ for $u$ and $v$,
performing a linear stability analysis of the disease-free state,
we find that the system defined by Eqs.~\eqref{eq:uv} exhibits traveling
wave solutions with front velocity given by\begin{equation}
c=\frac{2\alpha\omega^{+}\sqrt{D\left(2+\frac{\omega^{-}}{\omega^{+}}\right)}}{\alpha+\omega^{-}+2\omega^{+}}.\label{eq:MV}\end{equation}
If the forward and return rates $\omega^{+}$ and $\omega^{-}$ are
significantly different, two extreme cases can be considered. In the
limit $\omega^{+}\rightarrow0$ we find $c\rightarrow0$ as expected,
i.e. no propagation can be sustained in the limit of individuals not
leaving their base. If the backward rate $\omega^{-}$ is small, the
system is determined exclusively by the forward rate $\omega^{+}$. 

It is instructive to first consider a balanced system, i.e. $\omega^{+}=\omega^{-}=\omega$.
In this case Eq.~(\ref{eq:MV}) simplifies to \emph{\begin{equation}
c=\frac{2\sqrt{6D}\alpha\omega}{\alpha+3\omega}.\label{eq:BDV}\end{equation}
}The front velocity as a function of infection rate $\alpha$ and
travel rate $\omega$ as well as results of stochastic numerical simulations
are depicted in Fig.~\ref{fig:frontspeeds}. For comparison, the
front velocity of the reaction-diffusion scenario with the same global
travel rate $\omega$ is depicted as well. The velocity in this case
is given by Eq.~\eqref{eq:c_FK} which increases with travel rate
according to $\sim\sqrt{\omega}$. In contrast, the natural mobility
model exhibits a saturation of the front velocity with increasing
travel rate. From Eq.~\eqref{eq:BDV} it follows that the asymptotic
value of the velocity is proportional to the reaction rate $\lim_{\omega\rightarrow\infty}c=2\alpha\sqrt{2D/3}$. 

The existence of a saturation is a consequence of natural mobility
patterns that are restricted to individual mobility networks. Likewise
the unbounded increase in front velocity in a reaction-diffusion systems
is a consequence of the unnatural assumption that increasing the travel
rate also increases an individual's access to the entire system. In
the more realistic natural mobility model increasing $\omega$ only
increases the rate of travel between the base and the two neighboring
sites and does not imply faster coverage of the entire system.

\begin{figure}[t]
\hfill{}\includegraphics[width=0.49\textwidth]{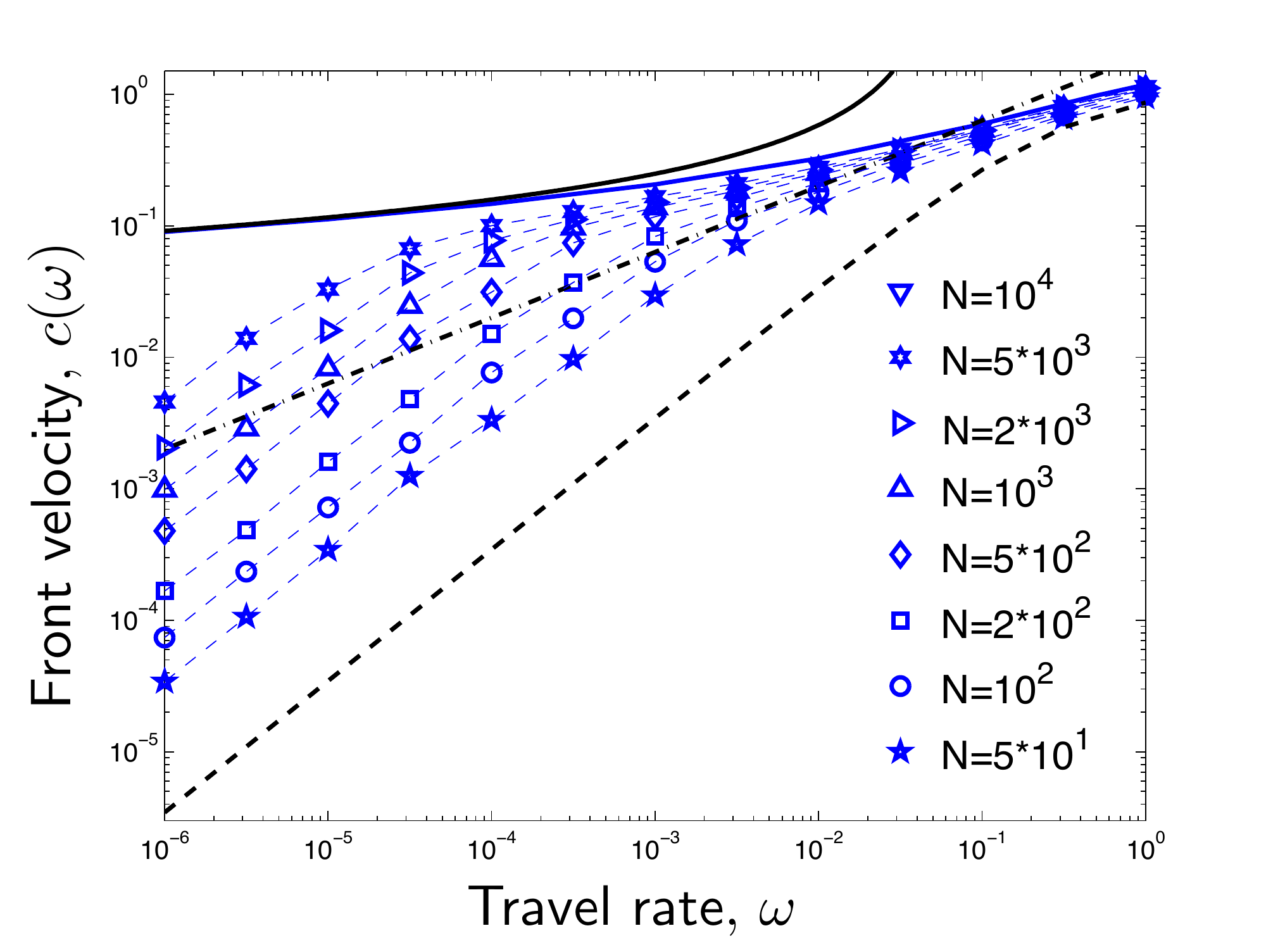}\hfill{}

\caption{\textbf{Front speed in the limit of small travel rates.} \emph{Bottom
Left:} Front velocity $c(\omega)$ in the regime of low travel rates
$\omega$. Symbols reflect stochastic simulations for different values
of particles per site. Dashed and dash-dotted lines represent analytical
results given by Eqs.~\eqref{eq:MV} and~\eqref{eq:c_FK}, respectively.
Solid blue line corresponds to the numerical solution of Eqs.~\eqref{eq:Bidi}
and the solid black line represents the analytical result of Eq.~\eqref{eq:c_wsmall}.
\label{fig:lowomega}}

\end{figure}

In contrast to the reaction-diffusion system with only one rate parameter
$\omega$, the natural human mobility model possesses two travel rates,
$\omega^{+}$and $\omega^{-}$ . The total flux between two neighboring
locations is given by $F_{nm}=\omega^{+}N_{n}^{n}+\omega^{-}N_{m}^{n}$,
for $m=n\pm1$. In equilibrium, detailed flux balance requires $F_{nm}=F_{mn}.$
Comparing to the reaction-diffusion system, the total flux is given
by $F_{nm}^{0}=\omega N$. In order to compare the dynamics of both
systems quantitatively it is plausible to gauge both systems such
that flux is identical, i.e.

\begin{equation}
\omega^{+}N_{n}^{n}+\omega^{-}N_{m}^{n}=\omega N.\label{eq:fluxequaltiy}\end{equation}
To simplify the analysis it is convenient to introduce a commuting
ratio $\varepsilon=\omega^{+}/\omega^{-}$. In a situation in which
individuals dwell at their base location most of the time, the commuting
ratio is a small, $\varepsilon\ll1$. We can express forward and return
rates in terms of the global travel rate $\omega$ and the commuting
ratio $\varepsilon$ according to: \begin{equation}
\omega^{+}=\frac{1+2\varepsilon}{2}\omega\quad\text{and}\quad\omega^{-}=\frac{1+2\varepsilon}{2\varepsilon}\omega.\label{eq:omegas}\end{equation}
With these definitions expression~\eqref{eq:MV} can be rewritten
as\begin{equation}
c=\frac{2\alpha\varepsilon\sqrt{2D\left(2+1/\varepsilon\right)}}{1+2\varepsilon+2\alpha\varepsilon(1+2\varepsilon)^{-1}\omega^{-1}}.\label{eq:Vwae}\end{equation}
Fig.~\ref{fig:lowomega} depicts the front velocity as a function
of the commuting ratio at the fixed travel rate ($\omega=1$) and
illustrates a significant change of $c$ as a function of $\varepsilon$.
Since the commuting ratio is not defined for ordinary reaction-diffusion
systems, this effect cannot be captured in these systems.

\begin{figure}[t]
\hfill{}\includegraphics[width=0.49\columnwidth]{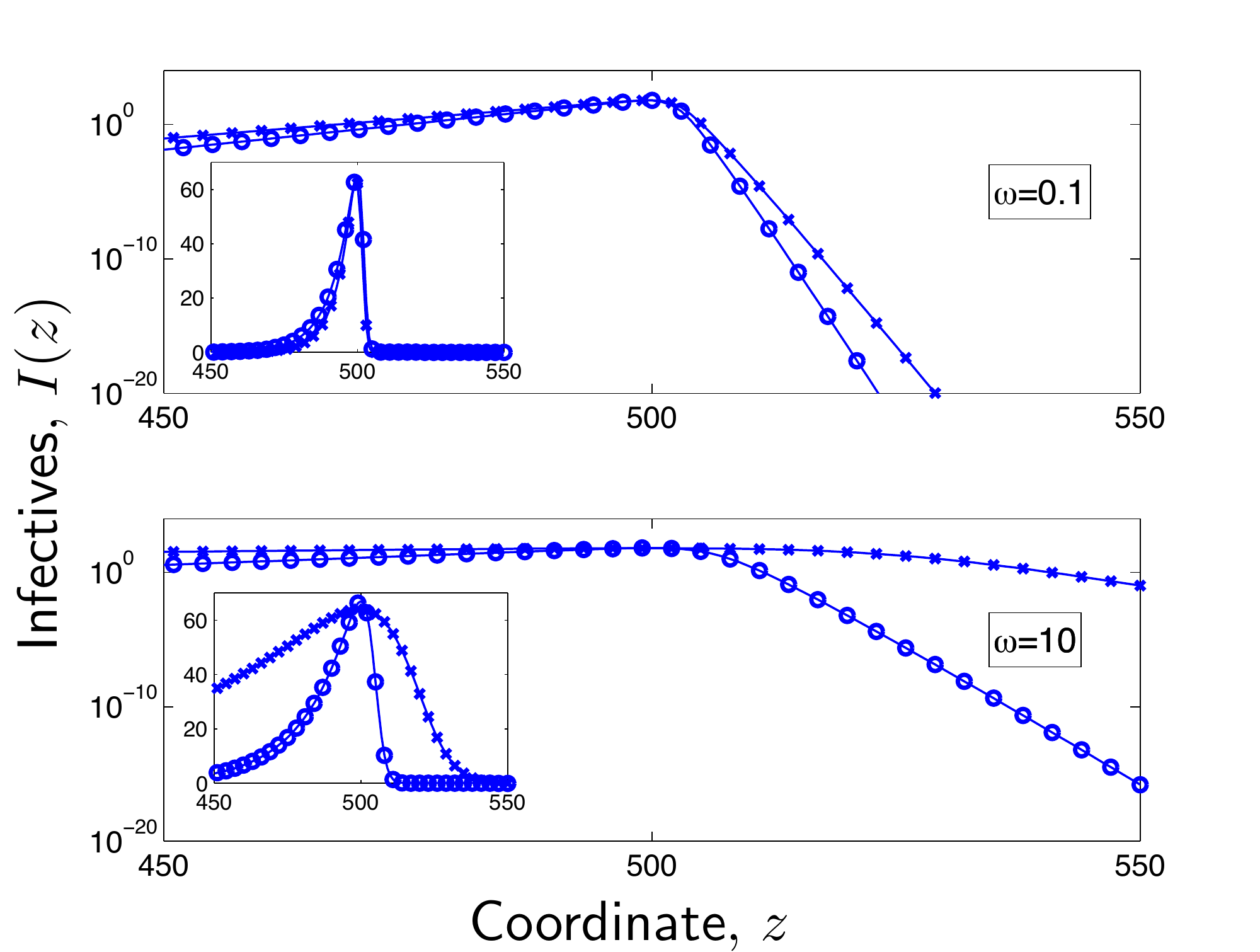}\hfill{}

\caption{\textbf{Front shape at the leading edge:} Front shape for different
global travel rates $\omega$ in the reaction-diffusion system (crosses)
and natural mobility model (circles) combined with local SIR dynamics.\label{fig:frontshape}}

\end{figure}

\begin{figure}[t]
\hfill{}\includegraphics[height=0.3\textheight]{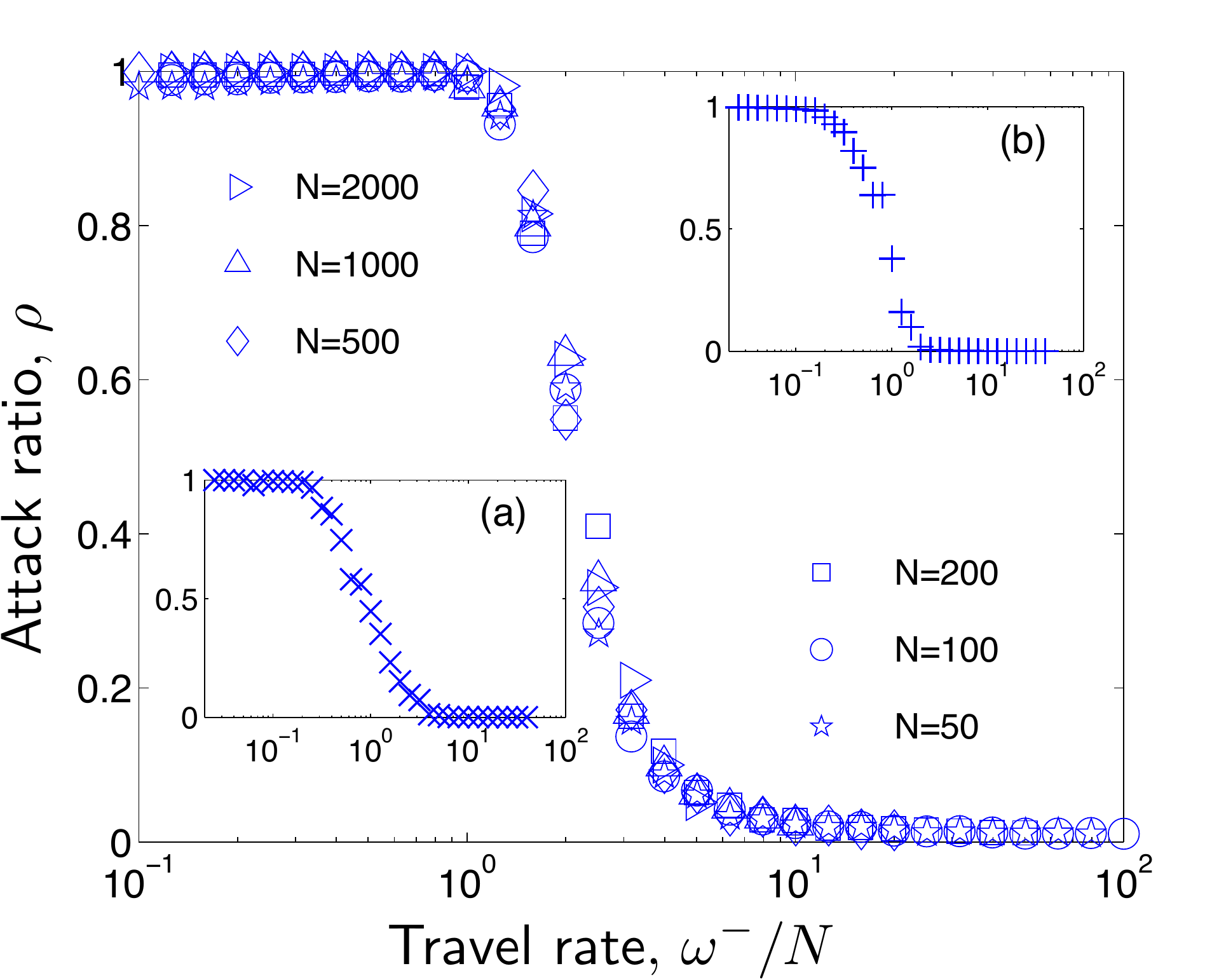}\hfill{}\includegraphics[height=0.3\textheight]{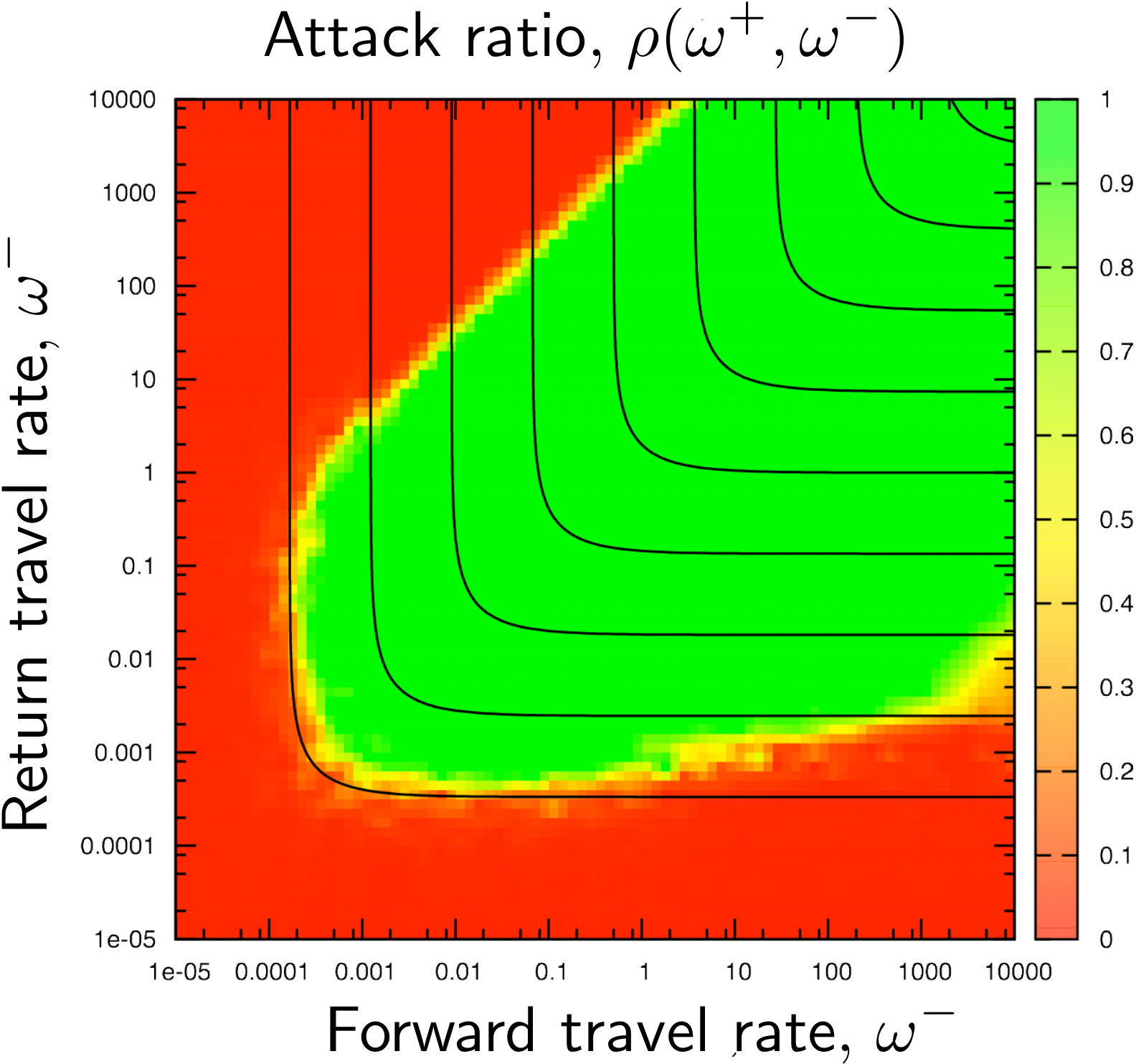}\hfill{}

\caption{\textbf{A novel type of invasion threshold.} \emph{Left:} The panel
depicts the results of stochastic simulations of an SIR epidemic in
combination with natural mobility patterns. The attack ratio $\rho(\omega^{-})$
as a function of the return travel rate $\omega^{-}$ exhibits a critical
transition and vanishes for return rates exceeding a critical value.
The function $\rho(\omega^{-})$ is shown for a 1-d lattices system
with $100$ sites with $N$ agents/site. \emph{Insets:} (a) $\rho(\omega^{-})$
for a scale-free network ($\gamma=1.5$) of $10^{3}$ nodes populated
uniformly with $\langle N\rangle=250$, $k_{\text{min}}=5$ and $k_{\text{max}}=50$;
(b) $\rho(\omega^{-})$ for an Erd\H{o}s-R\'enyi network with 500
nodes and $\langle k\rangle=10$. Results were averaged over 50 realizations.
Global travel rate was kept constant at $\omega=1$. Epidemic parameters
are $\alpha=1$, $\beta=0.1$. \emph{Right:} The attack ration as
a function of both travel rate parameters $\omega^{\pm}$ for a lattice
system. Solid lines represent constant global rate $\omega$ curves
with logarithmic spacing. The total flux increases from bottom left
to top right. Green regions corresponds to sustained outbreaks, red
reflects the extinction regime. Epidemic rates, lattice and averaging
parameters are the same as in the left panel.\label{fig:threshold1}}

\end{figure}

A surprising result in the behavior of the front velocity for very
low travel rates is depicted in Fig.~\ref{fig:frontspeeds} which
illustrates a noticeable deviation of the stochastic system from both,
the analytical prediction of Eq.~\eqref{eq:MV} and the results of
the numerical solution to Eqs.~\eqref{eq:Bidi}. The deviation of
the numerical solution from the Monte Carlo simulations for small
$\omega$ is due to the finite number of agents per site. We observe
a crossover from a linear scaling with $\omega$ (symbols) towards
the numerical mean-field solution (solid blue line). The regime of
low travel rates effectively corresponds to high infection rates ($\alpha\gg\omega$).
This implies that an outbreak takes place almost instantaneously in
a neighboring location and and epidemic essentially jumps from one
location to the next. The rate of hopping is proportional to $N\omega$,
i.e. to the flux of individuals between locations, where $N$ is a
typical of individuals per site. The crossover from discrete to continuous
behavior occurs when $\alpha\sim\omega N$, i.e. $\omega_{\text{c}}\sim\alpha/N$.
Note that the slow convergence of the velocity towards zero with decreasing
travel rate can be understood qualitatively: Let's consider just two
locations with agents that can travel between them. Without loss of
generality we consider an SI epidemic. At the beginning of the epidemic
the number of infecteds in the second location is small and we can
linearize the standart SI dynamics for the second population, i.e.
\begin{equation}
\frac{dj_{2}}{dt}\approx\alpha j_{2}+\omega j_{1},\label{eq:linearbla}\end{equation}
where we neglect the backward flux of the individuals from the second
location. Integrating Eq.~\ref{eq:linearbla} by means of the integrating
factor $j_{2}(t)=e^{\alpha t}\int_{0}^{t}d\tau j_{1}(\tau)$ and using
the solution of the SI model for the first location $j_{1}(t)=1/(1+ae^{-\alpha t})$
with $a=1-j_{1}(0)/j_{1}(0)$ yields\begin{equation}
j_{2}(t)=e^{\alpha t}\omega\left(\ln\frac{1+ae^{-\alpha t}}{1+a}+\ln e^{\alpha t}\right).\label{eq:j2}\end{equation}
As $t\rightarrow\infty$ we have $\ln(1+ae^{-\alpha t})\approx0$,
and thus $j_{2}(t)\sim\omega e^{\alpha t}$. The time lag between
outbreaks in both populations is given by $\Delta\tau(q)=\tau_{2}(q)-\tau_{1}(q)$,
where $\tau_{1}$ and $\tau_{2}$ are times when the concentrations
of infecteds attains some threshold $q$ in the first and second location
respectively. It follows that\begin{equation}
\Delta\tau(q)\sim\ln\frac{q}{\alpha\omega}-\ln\frac{q}{1-q}-\ln a,\label{eq:dtq}\end{equation}
and for $q=1/2$ we obtain for the velocity $c\sim\Delta\tau^{-1}$
and thus\begin{equation}
c\sim\left(-\ln\frac{\alpha\omega}{2}-\ln a\right)^{-1}.\label{eq:c_wsmall}\end{equation}
This expression reproduces the slow convergence of the front velocity
towards zero with decreasing travel rate which is in agreement with
the results obtained by numerical solution of the mean-field problem.

\subsection{Front shape}

Front velocity is tightly connected with the shape of the front at
the leading edge. In Fig.~\ref{fig:frontspeeds}~ front shapes as
obtained from numerical solutions for both reaction-diffusion and
natural mobility SIR models are shown. We observe that the reaction-diffusion
slope is much more strongly affected by the global travel rate $\omega$.
This is in contrast to the system with natural mobility patterns in
which, just like the front velocity, the front shape converges to
a fixed state as $\omega$ increases. The particular front shape is
related to Kendall waves that have been empirically observed~\cite{Post2}.

\subsection{Invasion Thresholds in natural human mobility models}

In addition to questions concerning the velocity of disease propagation,
it is of fundamental importance to assess the conditions under which
an epidemic propagates at all. Usually such a condition takes the
form of a threshold in a system parameter. Thea most prominent example
is the basic reproduction number, given by $R_{0}=\alpha/\beta$ for
a SIR model. It quantifies the a number of secondary cases caused
by a single infected individual in a totally susceptible population~\cite{may}.
If $R_{0}>1$ an outbreak occurs, otherwise the epidemic wanes. Another
threshold parameter in the metapopulation reaction-diffusion framework
is the global invasion threshold. It represents the minimal required
flux of individuals traveling between two locations~\cite{Vesp1}
in order for a disease to propagate spatially. 

\begin{figure}[t]
\hfill{}\includegraphics[height=0.25\textheight]{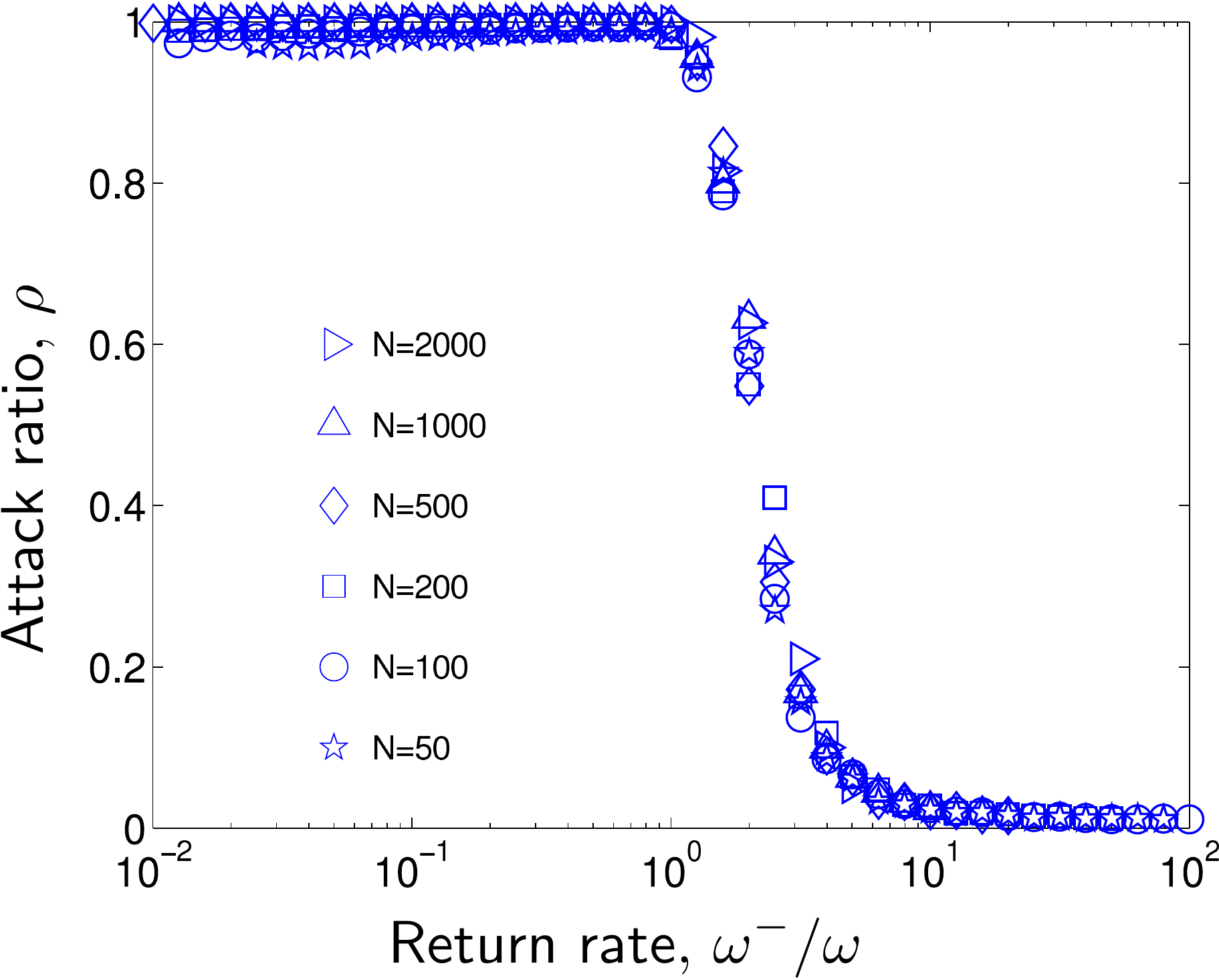}\hfill{}\includegraphics[height=0.25\textheight]{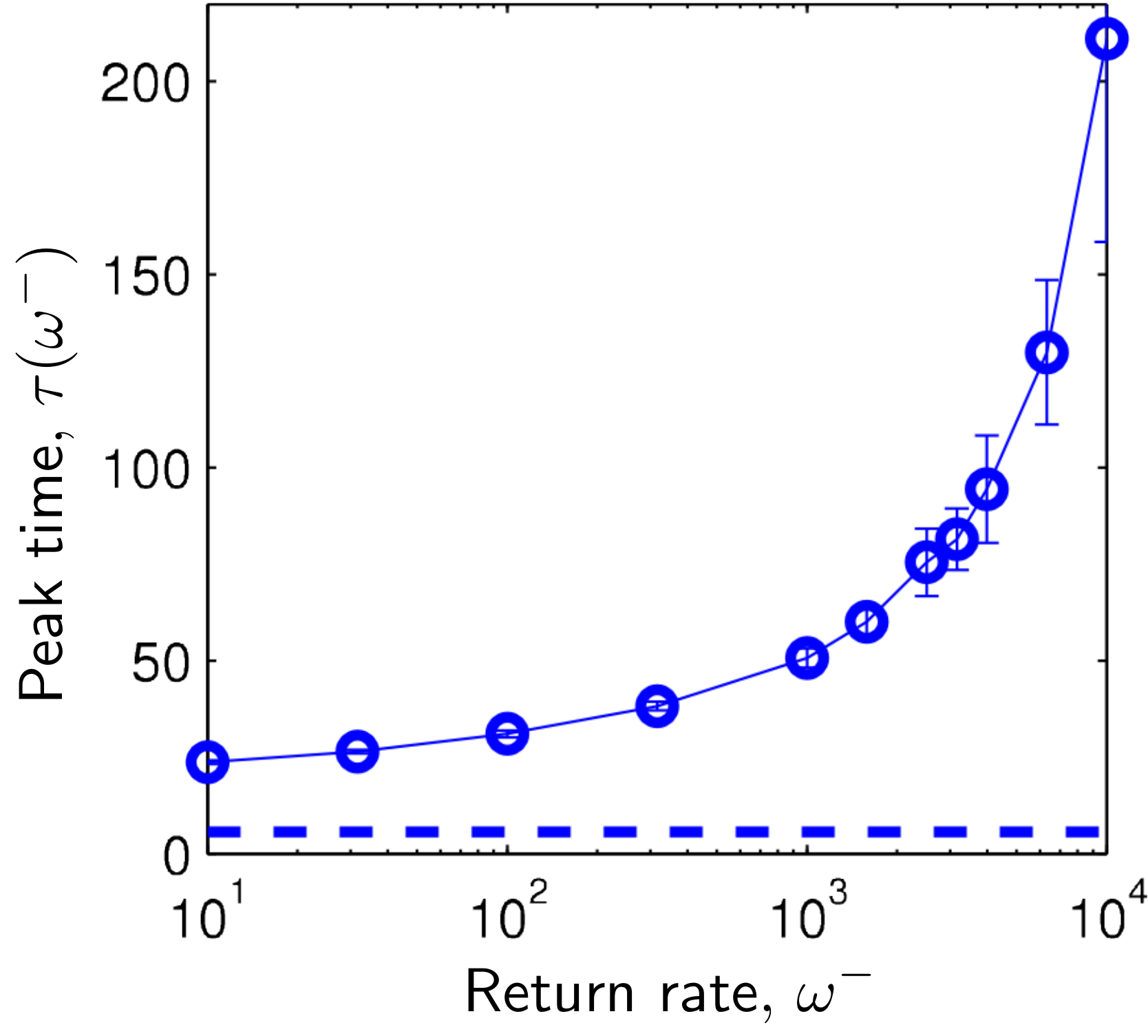}\hfill{}

\caption{\textbf{Effects in stochastic system:} \emph{Left:} The behavior of
the attack ratio $\rho$ as a function of the ratio of return rate
and $\omega^{-}$ global rate $\omega$. Other parameters are identical
to those of the 1-d system in Fig.~\eqref{fig:threshold1}. \emph{Right:}
Time lag $\tau(\omega^{-})$ as a function of $\omega^{-}$ for the
natural mobility model (circles) and ordinary reaction-diffusion (dashed
line) on a scale-free network (with scaling exponent $\gamma=1.5$
and $10^{3}$ nodes with an average of $\langle N\rangle=250$ individuals
per site. The curves depiced averages over 50 stochastic realizations.\label{fig:thresh2}}

\end{figure}

One of the most striking properties of the natural mobility model
of Eqs.~(\ref{eq:Bidi}) is the existence of a novel type of threshold
that is only determined by the return rate $\omega^{-}$, or equivalently
by the typical time an individual spends at a distant location. The
existence of this threshold is evident from Fig.~\ref{fig:threshold1}~
that depicts the attack ratio $\rho$ (the total fraction of infecteds
during an epidemic) as a function of the return rate $\omega^{-}$
on a) a one-dimensional lattice, b) and Erd\H{o}s-R\'enyi network
and c) an uncorrelated scale-free network. For low return rates the
attack ratio is close to unity, as expected, a global outbreak occurs.
However, with growing values of the return rate, the attack ratio
drops almost to zero, i.e. no global outbreak occurs. The regime of
high return rates corresponds to small dwelling times at distant locations.
This implies that an infected does not have sufficient time to transfer
the disease to susceptibles in unaffected locations before returning
home. This effect is absent in reaction-diffusion systems. This novel
type of threshold is a direct consequence of the properties of natural
human mobility.

To assess the mutual impact of all travel parameters, i.e. the forward
and return travel rates as well as the total flux on the dynamics
we calculated the attack ratio for various parameter values on a homogeneous
lattice. The results are presented in Fig.~\ref{fig:threshold1}.
Note that our model exhibits the global $\omega$-limited invasion
threshold that is also present in ordinary reaction-diffusion systems.
However, for large total flux $\omega$, the system exhibits a global
outbreak only if the return rate $\omega^{-}$ is sufficiently small.
Increasing the return rate the system enters a region that lacks a
global outbreak. Consequently, only the return rates $\omega^{-}$
is a limiting factor. This novel threshold can be estimated analytically.
In the same spirit as introduced recently~\cite{CollizaVespignani2007,Belik:2011}
we find the following threshold relation:\begin{equation}
\frac{N\omega}{\beta+\omega^{-}}(R_{0}-1)>1.\label{eq:2}\end{equation}
The inverse of the sum of return rate $\omega^{-}$ and recovery rate
$\beta$ provides the typical time an infected individual spends on
a distant location in the infected status. Using the relation $\omega=2\omega^{+}\omega^{-}/(2\omega^{+}+\omega^{-})$
(compare~\eqref{eq:omegas}), we can write explicitly\begin{equation}
\frac{\omega^{-}}{\beta}<\frac{\omega^{+}}{\beta}2N(R_{0}-1)-1.\label{eq:2exp}\end{equation}
Note, that from~\eqref{eq:2} the empirically observed scaling $\rho=\rho\left(N\omega/\omega^{-}\right)$
follows, see Fig.~\ref{fig:threshold1} and ~\ref{fig:thresh2}.
The figures exhibit the expected collapse of the data according to
this scaling. The pronounced difference between ordinary reaction-diffusion
systems and the natural mobility model is also captured in Fig.~\ref{fig:thresh2}
that illustrates the epidemic peak time $\tau=\int I(t)tdt/\int I(t)dt$
on the return rate. The figure shows that by varying the return rate
one can substantially increase the peak time in the natural mobility
model, in contrast to the reaction-diffusion model.

\section{Discussion}

\begin{figure}[t]
\includegraphics[width=1\textwidth]{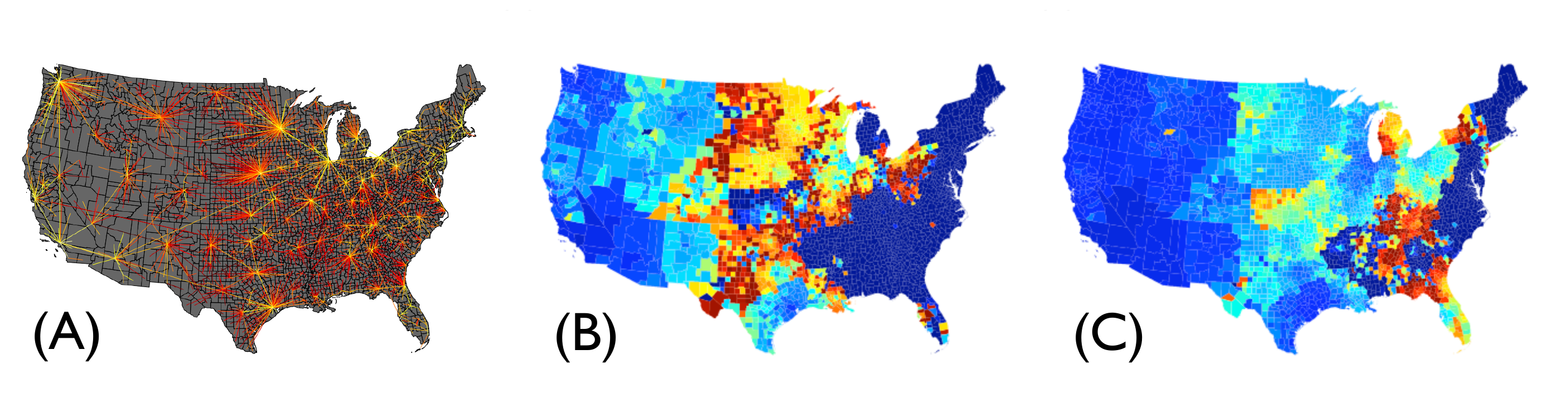}\caption{\textbf{Natural human mobility and disease dynamics in realistic scenarios.}
(A): The minimal spanning tree of a multiscale mobility network in
the United States. This network represents the skeleton of the most
important transportation routes in a more complete network~\cite{Thiemann:2010vb,DBNature}.
(B), (C): Comparison of the evolution of the SIR epidemic with initial
outbreak in Los Angeles for the natural mobility model (B) and ordinary
reaction-diffusion model (C). Snapshot of the number of infecteds
after approximately 4 weeks. Color encode from high (red) to low (blue)
the relative number of infecteds per county. \label{fig. 6}}

\end{figure}

An unprecedented amount of information on human mobility available
today requires adequate models that correctly capture the key features
of natural human mobility in order to understand, describe and predict
the dynamics of human mediated contagion phenomena. In the present
work we pursued this goal and formulated an approach that can account
for important features of natural human mobility that are intuitive
and observed empirically. The model is based on a metapopulation approach
assuming well-mixed local populations and explicit incorporation of
individual mobility patterns conditional on base or home location.
We considered regular bi-directional movements of the host between
base locations and accessible destinations, systematically analyzed
the model and compared it to established modeling approaches, e.g.
effective force of infection and reaction-diffusion systems. We demonstrated
that the latter are limiting cases of the natural human mobility model
at low and high travel rates, respectively. For a regular lattice
we derived a generalization of the Fisher-Kolmogorov equation and
found that contrary to the reaction-diffusion approach, the front
velocity of the epidemic does not increase unboundedly with increasing
global travel rate, but saturates at a maximum level.

Although results for lattice and artificial random network topologies
are extremely helpful in gaining fundamental insight into the dynamics
and consequences of natural human mobility patterns on the patterns
of disease spread, they at best mimic real world scenarios. A future
task will be to investigate to what level these novel effects prevail
in more realistic settings, i.e. real world mobility networks on which
individuals move and transmit disease. We hypothesize that effects
that are so dominant in parsimonious systems of the type described
by the model of Eqs.~\ref{eq:Bidi} will also be present in more
complex settings. Evidence for this has recently been revealed in
a multiscale metapopulation system~\cite{Balcan:2011gv}. In order
to illustrate the pronounced difference in disease dynamic patterns
that are generated by ordinary reaction diffusion models on one hand
and natural human mobility models on the other, we simulated both
systems on the backbone of a realistic, multiscale mobility network
in a real geographic setting. Fig.~\ref{fig. 6} illustrates snapshots
of the timecourse of disease spread generated by both models. Nodes
in the network are approx. 3000 counties in continental United States
and links between them resemble the traffic flux. We observe a significantly
smaller spreading speed in the natural mobility scenario compared
with the reaction-diffusion model. This implies that estimates of
spreading speeds, also in these more realistic setting, could have
been overestimated in the past by models that rely on ordinary diffusion
as a dispersal mechanism.

Although the study of natural human mobility on disease dynamics and
related human mediated contagion processes requires more attention
in future investigations, the results presented here as well as in
our previous work~\cite{Belik:2009fi,Belik:2011} will serve as a
useful guide for developing more reliable large scale computational
models for disease dynamics. The substantial differences of ordinary
reaction-diffusion dynamics and the novel type of natural human mobility
disperal suggest that disperal mechanism are among the most important
modeling ingredients that require particular care when implemented
in large scale computational models that are designed to make quantitative
forecasts. 

D.B. and V.B. acknowledge support from the Volkswagen Foundation.
D.B. acknowledge support from EU-FP7 grant Epiwork. 

\bibliographystyle{epj}
\bibliography{biblio}

\begin{thebibliography}{24}

\bibitem{may}
R.M. Anderson, R.M. May, \emph{Infectious Diseases of Humans} (Oxford
  University Press, 1991)

\bibitem{HBGPnas}
L.~Hufnagel, D.~Brockmann, T.~Geisel, Proc. Natl. Acad. Sci. USA \textbf{101},
  15124 (2004)

\bibitem{Fraser:2009ek}
C.~Fraser, C.A. Donnelly, S.~Cauchemez, W.P. Hanage, M.D. Van~Kerkhove, T.D.
  Hollingsworth, J.~Griffin, R.F. Baggaley, H.E. Jenkins, E.J. Lyons et~al.,
  Science \textbf{324}(5934), 1557 (2009)

\bibitem{DBNature}
D.~Brockmann, L.~Hufnagel, T.~Geisel, Nature (London) \textbf{439}, 462 (2006)

\bibitem{Gonzalez2008}
M.C. Gonz{\'a}lez, C.A. Hidalgo, A.L. Barab{\'a}si, Nature (London)
  \textbf{453}, 779 (2008)

\bibitem{Rileyreview}
S.~Riley, Science \textbf{316}, 1298 (2007)

\bibitem{Brockmann2007}
D.~Brockmann, L.~Hufnagel, Phys. Rev. Lett. \textbf{98}, 178301 (2007)

\bibitem{kpp}
A.~Kolmogorov, I.~Petrovsky, N.~Piscounov, Bull. de l'univ. d'{\'e}tat {\`a}
  Moscou, S{\'e}r. internat., sect. A \textbf{1}, 1 (1937)

\bibitem{fisher}
R.A. Fisher, Ann. Eugenics \textbf{7}, 355 (1937)

\bibitem{rvachev}
L.A. Rvachev, I.M. Longini, Math. Biosc. \textbf{75}, 3 (1985)

\bibitem{Vesp1}
V.~Colizza, R.~Pastor-Satorras, A.~Vespignani, Nature Physics \textbf{3}(4),
  276 (2007)

\bibitem{Rushton1955}
S.~Rushton, A.J. Mautner, Biometrika \textbf{42}, 126 (1955)

\bibitem{hagernaas}
T.J. Hagernaas, C.A. Donnelly, N.M. Ferguson, J. Theor. Biol. \textbf{229}, 349
  (2004)

\bibitem{Belik:2009fi}
V.V. Belik, T.~Geisel, D.~Brockmann, in \emph{Proceedings of 2009 International
  Conference on Computational Science and Engineering} (IEEE, 2009), pp.
  932--935, ISBN 978-1-4244-5334-4

\bibitem{Brockmann:2009tf}
D.~Brockmann, \emph{{Human Mobility and Spatial Disease Dynamics}}, in
  \emph{Reviews of Nonlinear Dynamics and Complexity} (Wiley-VCH, 2009), pp.
  1--24

\bibitem{Chaoming-Song-and-Zehui-Qu-and-Nicholas-Blumm-and-Albert-Laszlo-Barabasi-:2010la}
C.~Song, Z.~Qu, N.~Blumm, A.L. Barab{\'a}si, Science \textbf{327}, 1018 (2010)

\bibitem{Song:2010kx}
C.~Song, T.~Koren, P.~Wang et~al., Nat. Phys. \textbf{6}, 818 (2010)

\bibitem{Belik:2011}
V.~Belik, T.~Geisel, D.~Brockmann, Phys. Rev. X  (2011), in press.

\bibitem{Balcan:2011gv}
D.~Balcan, A.~Vespignani, Nature Physics  (2011)

\bibitem{bidirect1}
L.~Sattenspiel, K.~Dietz, Math. Biosc. \textbf{128}, 71 (1995)

\bibitem{Keeling2002}
M.J. Keeling, P.~Rohani, Ecol. Lett. \textbf{5}, 20 (2005)

\bibitem{CollizaVespignani2007}
V.~Colizza, A.~Vespignani, Phys. Rev. Lett. \textbf{99}, 148701 (2007)

\bibitem{Post2}
U.~Naether, E.B. Postnikov, I.M. Sokolov, Eur. Phys. J. B \textbf{65}, 353
  (2008)

\bibitem{Thiemann:2010vb}
C.~Thiemann, F.~Theis, D.~Grady, R.~Brune, D.~Brockmann, PLoS ONE
  \textbf{5}(11), e15422 (2010)

\end{thebibliography}

\end{document}